# Gamow's alpha-decay theory revisited

K. D. Krori[1] and Samrat Dey[2, 3]

[1]Theoretical Physics Foundation, Lachitnagar, Bylane – 7, Sublane – 2, Guwahati – 781007 (India)
[2]Physics Department, Assam Don Bosco University, Sonapur – 782402 (India)
[3]Physics Department, Pragjyotish College, Guwahati – 781009 (India)

E-mail: samratdgr8@gmail.com

Abstract

G. Gamow's alpha-decay theory (1928), although successful, has a wrong phenomenological argument that an alpha-particle inside a radioactive nucleus moves back and forth through the dense mass of nucleons (retaining its identity) a number of times before it comes out. This short paper seeks to get over this shortcoming by deriving principally the Gamow's theory through a slightly alternative approach by avoiding the problematic touching frequency in the original theory.

## 1. Introduction

For nearly a century, it has been taught in undergraduate level quantum mechanics that the alpha emission from a radioactive nucleus occurs by quantum tunnelling through a potential barrier after the particle moves back and forth a number of times inside the nucleus, as Gamow thought [1, 2]. An alpha-particle is a composite body consisting of two protons and two neutrons and a nucleus is an object composed of a compact mass of protons and neutrons. One may reasonably ask: is it possible for an alpha-particle to move from potential barrier to potential barrier a number of times through the dense crowd of nucleons inside a nucleus before emission retaining its identity? The theory is, thus, certainly flawed as the alpha particle should not exist inside the nucleus, not to say it's oscillating and touching of the surface [3, 4].

Despite this short coming, Gamow's theory is even now taught at the undergraduate level across the world in its original form [2, 4]; the reason being the following successes: (a) it catches the main physics of quantum penetrability and (b) it is really simple and (c) it is convenient to apply to study the alpha decay half-life once the Q value is known. The limited objective of this short paper is to present the theory in a physically corrected and rational form, and yet retain all the advantages mentioned above. We, thus, present an alternative treatment of the theory. We consider alpha decay as a form of nuclear disintegration into a big fragment (a daughter nucleus) and a small fragment (an alpha-particle) as the alpha-particle does not exist as such inside a nucleus. Certainly, the corrected theory is only a modification of the original macroscopic theory of Gamow and not a microscopic picture of the alpha formation inside the nucleus, just like the original Gamow's theory (there are many scholarly articles on sophisticated microscopic theories available in literature [5-8]). We achieve this with the introduction of a new equation by replacing the touching frequency of Gamow (which is problematic and wrong phenomenological argument by itself); resulting equation is very much similar to the original one of Gamow.

## 2. Nuclear disintegration and alpha-decay

In 1939, N. Bohr and J. A. Wheeler [9] for the first time studied disintegration of a radioactive nucleus highly agitated after being bombarded by a neutron. They likened such a nucleus to an internally agitated liquid drop [5]. Violent convulsions occur inside the agitated nucleus and rapid deformations follow. It may have a symmetrical deformation shown in Fig. 1(b) or an asymmetrical deformation shown in Fig. 2(b). In this state, if the long-range Coulomb force between A and B dominates over the short-range attractive nuclear force between them, a neck would appear between them [Fig. 1(c) and Fig. 2(c)] and they would end up in violent separation in opposite directions.

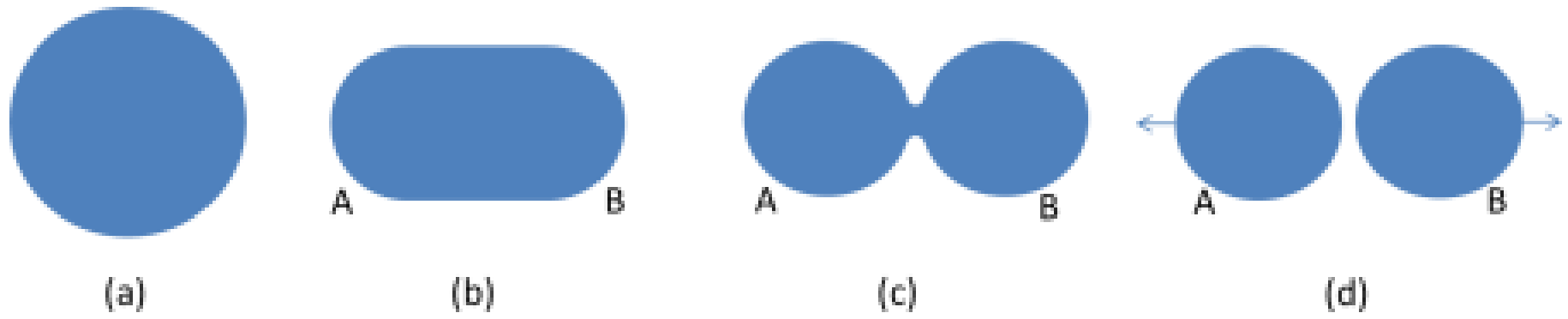

Fig. 1 Equal fragments in disintegration

However, in alpha-decay, the short-range attractive nuclear force dominates over the long-range repulsive Coulomb force. In this case quantum tunnelling plays the role of nuclear disintegration.

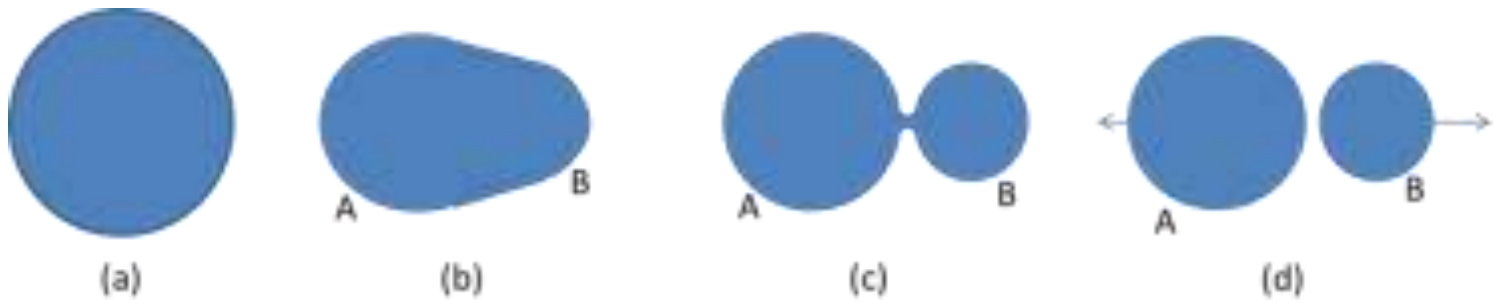

Fig. 2 Unequal fragments in disintegration

It is reasonable to think that a naturally agitated radioactive nucleus may undergo periodic deformation (Fig. 3). It may so happen that a bulge occurs on the right and then on the left and again on the right and then on the left and so on. $A_1B_1B_2A_2$ constitutes a cycle of deformation. $A_3B_3B_4A_4$ is the next cycle of deformation and so on. Each of the bulges, $B_1$, $B_2$, $B_3$, etc., represents an alpha-particle attempting to break through the potential barrier of the daughter nucleus radially. Ultimately in (n) the alpha-particle breaks through the potential barrier of the daughter nucleus. In (m) the alpha-particle is detached from the daughter nucleus resulting in complete disintegration of the nucleus.

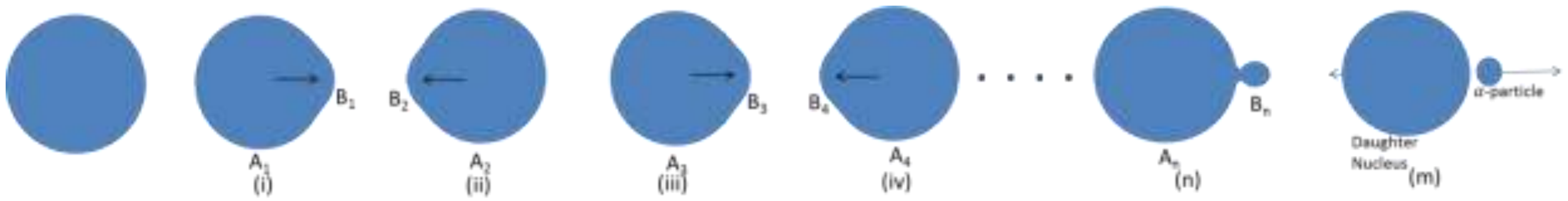

Fig. 3 Disintegration involving alpha-emission

Now we proceed to find the energy of the emitted alpha particle. To a good approximation we are guided by the idea that the frequency of the periodic movement of a micro-object multiplied by $h$ gives its energy. We therefore take $h\nu$ ($\nu$ being the frequency of periodic deformation of an isolated radioactive nucleus) as the energy of the nucleus. Then the emitted alpha particle will have the energy, $E$, given by

$$E = kh\nu \qquad \ldots(1)$$

where we have $0 < k \leq 1$.

## 3. Detailed calculations

Our mathematical treatment will have a fair degree of similarity to that of Gamow for obvious reasons, as discussed above. The alpha-decay is studied subject to the following assumptions:

(i) The nucleus has atomic number, $Z$, and retains *broadly* its spherical shape during periodic deformations.

(ii) The alpha-decay involves interaction between the alpha-particle of atomic number $Z_1 = 2$ and the daughter nucleus of atomic number, $Z_2 = Z - 2$.

(iii) The alpha-particle has radial motion while attempting to come out of the potential barrier (Fig. 4).

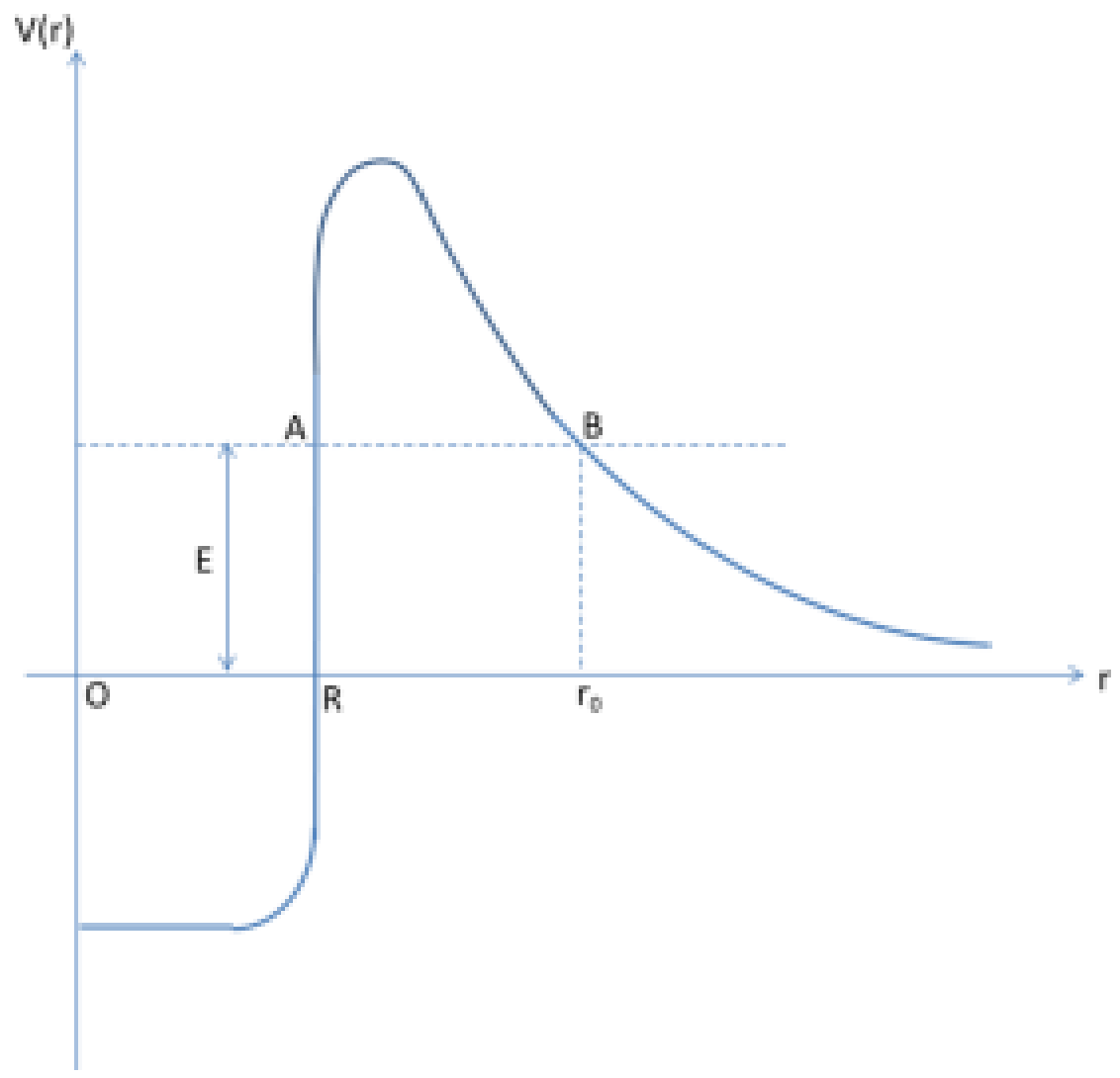

Fig. 4 Potential barrier in alpha-decay

(iv) Within the barrier, the alpha-particle is acted upon by the Coulomb force $= \frac{Z_1 Z_2 e^2}{r^2}$, $e$ being the charge of a proton.

(v) The alpha-particle with energy, $E$, emerges from the barrier at $r = r_0$.

On account of the continuity of $\psi$ (representaing the alpha-particle) at the boundaries at $r = R$ and $r = r_0$, we shall consider $\psi$ only within the barrier. The equation of motion is (for the radial motion of the alpha-particle)

$$\frac{d^2\psi}{dr^2} + \frac{2m_0}{\hbar^2}\left(E - \frac{Z_1 Z_2 e^2}{r}\right)\psi = 0 \qquad \ldots(2)$$

Writing $\psi = Ae^{y(r)/\hbar}$, we have from (2)

$$\hbar\frac{d^2y}{dr^2}+\left(\frac{dy}{dr}\right)^2-2m_0\left(\frac{Z_1Z_2e^2}{r}-E\right)=0 \qquad \text{...(3)}$$

Neglecting the first term which contains a very small factor, $\hbar$, we have

$$\left(\frac{dy}{dr}\right)^2=2m_0\left(\frac{Z_1Z_2e^2}{r}-E\right) \qquad (4)$$

Now, following Eq. (A6) (see Appendix and it may be noted that the disintegration is also quantum tunnelling process through the potential barrier), we obtain the fission probability

$$T=e^{-2\int_R^{r_0}\sqrt{\frac{2m_0}{\hbar^2}\left(\frac{Z_1Z_2e^2}{r}-E\right)}\,dr} \qquad \text{...(5)}$$

An alpha-particle approaching B from outside with energy, E, would be turned back. This means that

$$E=\frac{Z_1Z_2e^2}{r_0}$$

so that

$$r_0=\frac{Z_1Z_2e^2}{E} \qquad \text{....(6)}$$

Now the integral in Eq. (5) is evaluated.

$$\int_R^{r_0}\sqrt{\frac{2m_0}{\hbar^2}\left(\frac{Z_1Z_2e^2}{r}-E\right)}\,dr=\left(\frac{2m_0Z_1Z_2e^2}{\hbar^2}\right)^{\frac{1}{2}}\int_R^{r_0}\left(\frac{1}{r}-\frac{1}{r_0}\right)dr=\left(\frac{2m_0Z_1Z_2e^2}{\hbar^2}\right)^{\frac{1}{2}}\sqrt{r_0}\left[\cos^{-1}\left(\frac{R}{r_0}\right)^{\frac{1}{2}}-\left(\frac{R}{r_0}-\frac{R^2}{r_0^2}\right)^{\frac{1}{2}}\right] \qquad \text{...(7)}$$

At low energies (small $E$), $r_0$ is large according to Eq. (6) so that we make $r_0 \gg R$. Hence Eq. (7) reduces to

$$\int_R^{r_0}\sqrt{\frac{2m_0}{\hbar^2}\left(\frac{Z_1Z_2e^{\ 2}}{r}-E\right)}\,dr=\left(\sqrt{\frac{2m_0}{E}}\frac{Z_1Z_2e^2}{\hbar}\right)\left(\frac{\pi}{2}-\delta\right) \qquad \text{...(8)}$$

where $\delta=2\sqrt{\frac{R}{r_0}}=2\sqrt{\frac{R}{\frac{Z_1Z_2e^2}{E}}}$.

Using Eq. (8) in Eq. (5),

$$T=e^{-2\left(\sqrt{\frac{2m_0}{E}}\frac{Z_1Z_2e^2}{\hbar}\right)\left(\frac{\pi}{2}-\delta\right)} \qquad \text{...(9)}$$

Now, say, n alpha-like attempts are made before an alpha-particle breaks through the potential barrier, then obviously $T=\frac{C}{n}$, i.e., $n=\frac{C}{T}$, where $C$ is a constant and should have different values for different radioactive clusters. Again, the time between two such successive alpha-like attempts is $\frac{1}{2\nu}$. Then, the mean life of the decay process is, from Eq. (1), (with k=1),

$$\tau=n\times\frac{1}{2\nu}=\frac{C}{T}\frac{h}{2E}=Ce^{2\left(\sqrt{\frac{2m_0}{E}}\frac{Z_1Z_2e^2}{\hbar}\right)\left(\frac{\pi}{2}-\delta\right)}\times\frac{h}{2E} \qquad \text{...(10)}$$

The following is the expression for mean life of the alpha-decay process from Gamow's work [2] (C=1)

$$\tau=n\times\frac{1}{2\nu}=e^{2\left(\sqrt{\frac{2m_0}{E}}\frac{Z_1Z_2e^2}{\hbar}\right)\left(\frac{\pi}{2}-\delta\right)}\times 2R\sqrt{\frac{m_0}{2E}} \qquad \text{...(11)}$$

Now we take two examples for comparison between Eq. (10) and Eq. (11) as well as experimental values of their half-lives. .

Reaction 1: $Po_{84}^{212} \rightarrow Pb_{82}^{208} + \alpha(8.9\ MeV)$
Reaction 2: $U_{92}^{238} \rightarrow Th_{90}^{234} + \alpha(4.2\ MeV)$
From the values of $E$ and $Z_2$ we can calculate $\tau$ for these reactions corresponding to Eq. (10) and Eq. (11), remembering that $R = 1.07 \times 10^{-13} \times A^{1/3}\ units$; here we use cgs units. $\tau$, calculated by both the equations, is shown in Table 1.

Table 1: $\tau$ for two decay reactions with different $Z_2$

| Reaction | $A$ | $Z_2$ | $E$ (in MeV) | $\tau$ (in sec) [from Eq. (10)] | | $\tau$ (in sec) [from Eq. (11)] | Experimental $\tau$ (in sec) [10] |
|---|---|---|---|---|---|---|---|
| | | | | $C = 1$ | $C = 1.1 \times 10^{-5}$ | | |
| 1 | 212 | 82 | 8.9 | $1.36 \times 10^{-04}$ | $1.50 \times 10^{-09}$ | $3.62 \times 10^{-04}$ | $3 \times 10^{-07}$ |
| 2 | 238 | 90 | 4.2 | $6.83 \times 10^{22}$ | $7.51 \times 10^{17}$ | $12.99 \times 10^{22}$ | $1 \times 10^{17}$ |

Thus, we see that Eq. (10) with $C = 1$ and Eq. (11) lead to results which have same orders of magnitude. Nevertheless, both of these results are significantly different from their corresponding experimental value. However, this can be fairly addressed by choosing a suitable value of the parameter $C$.

We have so far considered that each bulge occurring in Fig. 3 represents an alpha-particle. But it may represent heavier clusters such as $^{14}$C, $^{20}$O, $^{24}$Ne, etc. also and we have what is popularly called cluster radioactivity [6, 7]. Eq. (10) can very well be applied to such phenomenon, as well.

## 4. Discussion

Our alternative approach has some positive aspects. In the first place, Gamow's idea of to-and-fro movement of an alpha-particle inside a compact nucleus before emission is reasonably averted. In the second place, it can be extended to study of cluster radioactivity at graduate level. Thirdly, there is no problem of preformation of emitted particles [8]. Finally, our results for $\tau$, from Eq. (10) with $C = 1$, appear to be similar to the original Gamow's results for $\tau$, from Eq. (11). This is encouraging for our endeavour thus far.

## References

1. Gamow, G. 1928 *Zeits f. Physik* ***51*** 204
2. Gasiorowicz, S. 1995 *Quantum Physics* 86-89 (John Wiley)
3. Krori, K. D. 2020 *Philosophical Strands in Physics* 73 (Theoretical Physics Foundation)
4. Krane, K. S. 1988 *Introductory Nuclear Physics* 251 (John Wiley & Sons)
5. Royer, G. 2000 *J. Phys. G.* **26** 1149-1170
(Our approach is different)
6. Lovas, R. G., Liotta, R.J., Insolia, A., Varga, K. and Delion, D.S. 1998 *Phys. Rep.* **294** 265-362.
7. Ismail, M., Ellithi, A. Y., Selim, M. M., Abou-Samra, N. and Mohamedien, O. A. 2020 *Phys. Scr.* **95** 075303

8. Zhang, H. F., Royer, G. and Li, J. Q. 2011 *Phys. Rev. C* **84** 027303.
9. Bohr, N. and Wheeler J. A. 1939 *Physical Review* ***56*** 426-450
10. Ghatak, A. and Lokanathan, S. 2004 *Quantum Mechanics: Theory and Applications* 455 (Springer Dordrecht)

# APPENDIX

(For ready reference)

A particle (represented by a function, $\psi$) of energy, $E$, in a potential field, V(x), (Fig. 5), enters a potential barrier at $x = a$ and passes out at $x = b$.

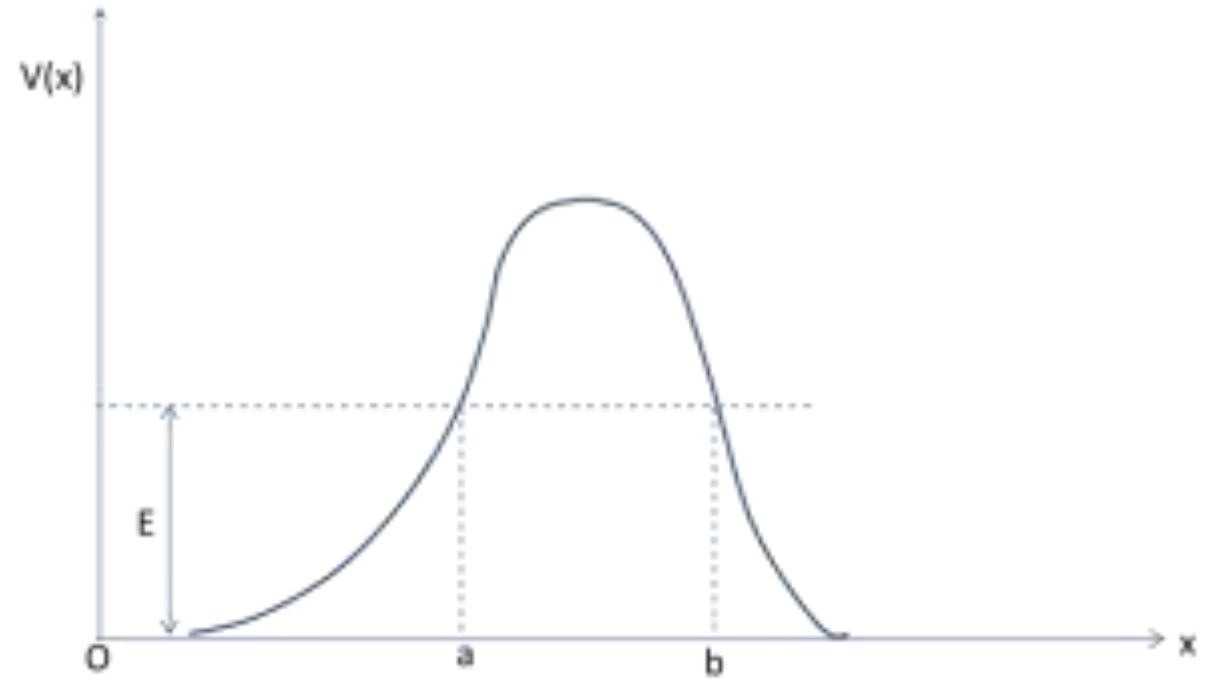

Fig. 5

The equation of motion is

$$\frac{d^2\psi}{dx^2} + \frac{2m_0}{\hbar^2}[E - V(x)]\psi = 0 \quad \text{...(A1)}$$

Writing

$$\psi = Ae^{y(x)/\hbar} \quad \text{...(A2)}$$

We obtain from (A1)

$$\hbar\frac{d^2y}{dx^2} + \left(\frac{dy}{dx}\right)^2 - F(x) = 0 \quad \text{...(A3)}$$

where $F(x) = 2m_0[V(x) - E]$. Neglecting the first term which contains a very small quantity, $\hbar$, $\frac{dy}{dx} = -\sqrt{F(x)}$, the minus sign is taken for a physical result, giving

$$y = -\int^x \sqrt{F(x)}dx + constant \quad \text{...(A4)}$$

On account of the property of continuity, $\psi$ has the same value across $x = a$ and $x = b$. Hence we have

$$\frac{\psi(b)}{\psi(a)} = e^{+\frac{1}{\hbar}[y(b)-y(a)]} = e^{-\frac{1}{\hbar}\int_a^b \sqrt{F(x)}dx} \quad \text{...(A5)}$$

Then the approximate form of the probability to break through the potential region is

$$T = \left|\frac{\psi(a)}{\psi(b)}\right|^2 = e^{-\frac{2}{\hbar}\int_a^b \sqrt{F(x)}dx} = e^{-\frac{2}{\hbar}\int_a^b \sqrt{2m_0[V(x)-E]}dx} \quad \text{...(A6)}$$

$T$ can be determined from a knowledge of $V(x)$.